\documentclass[12pt, tightenlines]{revtex4-1}

\usepackage{amsmath}
\usepackage{amssymb}
\usepackage{graphicx}
\usepackage{dcolumn}
\usepackage{bm}
\usepackage{color}

\begin{document}

\title{A Local Quantum Phase Transition  in YFe$_{2}$Al$_{10}$}
\author{W. J. Gannon}
\email{wgannon@physics.tamu.edu}
\affiliation{Department of Physics and Astronomy, Texas A \& M University, College Station, TX 77843-4242 USA}

\author{L. S. Wu}
\affiliation{Quantum Condensed Matter Division, Oak Ridge National Laboratory, Oak Ridge TN 37830-6477 USA}

\author{I. A. Zaliznyak}
\affiliation{Condensed Matter Physics and Materials Science Division, Brookhaven National Laboratory, Upton, NY 11973 USA}

\author{W. Xu}
\affiliation{Condensed Matter Physics and Materials Science Division, Brookhaven National Laboratory, Upton, NY 11973 USA}

\author{A. M. Tsvelik}
\affiliation{Condensed Matter Physics and Materials Science Division, Brookhaven National Laboratory, Upton, NY 11973 USA}

\author{J. A. Rodriguez-Rivera}
\affiliation{NIST Center for Neutron Research, National Institute of Standards and Technology, Gaithersburg, MD 20899 USA}

\author{Y. Qiu}
\affiliation{NIST Center for Neutron Research, National Institute of Standards and Technology, Gaithersburg, MD 20899 USA}

\author{M. C. Aronson}
\affiliation{Department of Physics and Astronomy, Texas A \& M University, College Station, TX 77843-4242 USA}


\begin{abstract}
A phase transition occurs when correlated regions of a new phase grow to span the system and the fluctuations within the correlated regions become long-lived. Here we present neutron scattering measurements showing that this conventional picture must be replaced by a new paradigm in \boldmath$\mathrm{YFe}_2\mathrm{Al}_{10}$, a compound that forms naturally very close to a \boldmath$T=0$ quantum phase transition. Fully quantum mechanical fluctuations of localized moments are found to diverge at low energies and temperatures, however the fluctuating moments are entirely without spatial correlations.  Zero temperature order in \boldmath$\mathrm{YFe}_2\mathrm{Al}_{10}$ is achieved by a new and entirely local type of quantum phase transition that may originate with the creation of the moments themselves.
\end{abstract}

\maketitle
Magnetic order arises from the growth of magnetic correlations, which become increasingly long-lived and extend over longer distances as the phase transition to magnetic order is approached. The magnetically ordered ground state can be destabilized by pressure, composition, or magnetic field, and there are extremal values of these nonthermal variables where order occurs only at $T=0$, the Quantum Critical Point (QCP). It is the strong quantum fluctuations associated with low dimensionality, or alternatively the frustration of competing interactions on lattices with certain geometries, that can suppress magnetically ordered phases to produce these QCPs. Magnetic order also requires magnetic moments, which in metals can be produced by different types of  $T=0$ instabilities. For spatially localized f-electrons, it is the Kondo compensation provided by  conduction electrons that determines whether a moment is retained at $T=0$. Mott physics governs the more delocalized d-electrons, where correlations among the mobile electrons may produce a spatially localized moment with a magnitude that can approach the large moments possible in insulators, or alternatively correlations so weak that they cannot induce even a tiny moment that could lead to magnetic order at a correspondingly low, but still nonzero, temperature.  Phase transitions leading to moment formation at $T=0$ are expected to have a very different character than those that lead only to magnetic order.

It has proven difficult to make a clean experimental distinction between QCPs that are related to magnetic order, involving a broken symmetry, and those that correspond to moment formation.  The conventional picture of classical phase transitions can be extended in certain systems to $T=0$, where  neutron scattering documents the growth of spatial and temporal correlations that are related to fluctuations of the order parameter~\cite{knafo2009,singh2011}. Only mean field behavior~\cite{hertz1976,millis1993,moriya1985} is observed, indicating that these systems lack strong quantum fluctuations. In contrast, neutron scattering experiments on ${\rm CeCu}_{6-x_{C}}{\rm Au}_{x_{C}}$
~\cite{schroeder2000} and ${\rm BaFe}_{1.85}{\rm Co}_{0.15}{\rm As}_{2}$~\cite{inosov2010,varma2015} find strong QC fluctuations and the breakdown of conventional Fermi liquid behavior near the wave vectors that will eventually become magnetic Bragg peaks in nearby AF phases. So far, there is no case where the comparison of experimental and theoretical QC phenomena definitively identifies QC fluctuations of a $T=0$ order parameter of any kind~\cite{sachdev2008}. Nonetheless, there is mounting  evidence that moment formation may play an important role near QCPs. In the Kondo breakdown scenario, proposed for f-electron heavy fermion compounds,  the QC fluctuations associated with magnetic order are strong enough to localize a moment-bearing electron~\cite{coleman2001,si2001}. The collapse of the Kondo effect may occur exactly at a magnetic QCP~\cite{paschen2004}, or simply close to one~\cite{friedemann2009,custers2010}. It is  accompanied not by order parameter fluctuations, as near a magnetic phase transition, by rather by QC fluctuations between two Fermi surfaces, one containing the electron that will be localized, and one that does not. A very different $T=0$ phase transition envisages moment formation as the consequence of a topological instability in a metal with strong electronic correlations~\cite{senthil2004,senthil2008}. Magnetic order plays no role, and so the correlations associated with this QCP are necessarily short-ranged, although the moment dynamics are definitively QC. Particularly appropriate for d-electron based metals, the orbital selective Mott transition (OSMT) provides a general theoretical structure~\cite{mvojta2010} for a phase transition where one or more orbitals can transition from being localized and magnetic, to delocalized and nonmagnetic~\cite{anisimov2002,demedici2005,demedici2009}.  Practically speaking, the emergence of a magnetic moment in a metal, either by a topological instability or by Mott physics, is very likely to lead to magnetic order, except in the most frustrated of systems.  Magnetic phase transitions at $T=0$  do not require simultaneous moment formation via electronic localization transitions, however we lack direct experimental evidence of the converse situation, where an electronic localization transition leading to moment formation can exist independently of magnetic order. It is significant that the neutron scattering results reported here show YFe$_2$Al$_{10}$ may be the first example of a metal on the verge of moment formation, possibly via an OSMT, but without any vestige of magnetic order~\cite{park2011}.

In materials that are magnetically ordered, or nearly so, magnetic correlations depend strongly on wave vectors $\bf{q}$ that reflect the spatial periodicity of the magnetic structure. Our inelastic neutron scattering measurements show (Fig. 1A,B) that the magnetic fluctuations in  YFe$_2$Al$_{10}$~ are very different.  Here,  the scattered intensity $I(q)$ is dominated by a broad ridge of scattering along wave vectors
$q$ parallel to [00L], lying in the critical $ac$ plane defined by the Fe-layers (inset, Fig. 1D)~\cite{thiede1998,kerkau2012,park2011}. Consistent with the $T/B^{0.6}$ scaling observed in the magnetization and specific heat~\cite{wu2014}, the scattering is strongly suppressed by magnetic fields $B$ (Fig. 1A). The critical part of $I(q)$ can be exposed by using similar data obtained at $9~T$ (Fig. 1A, right) as an improvised background for the $B=0$ data (Fig. 1A, left). Fig. 1B shows that the result is a weak and broad modulation of the field - dependent component of the scattering in the [0K0] direction $I(q_{K}$), perpendicular to the Fe-layers, with a breadth that extends over more than the full Brillioun zone.

\begin{figure}
\centering
\includegraphics[width=.8\linewidth]{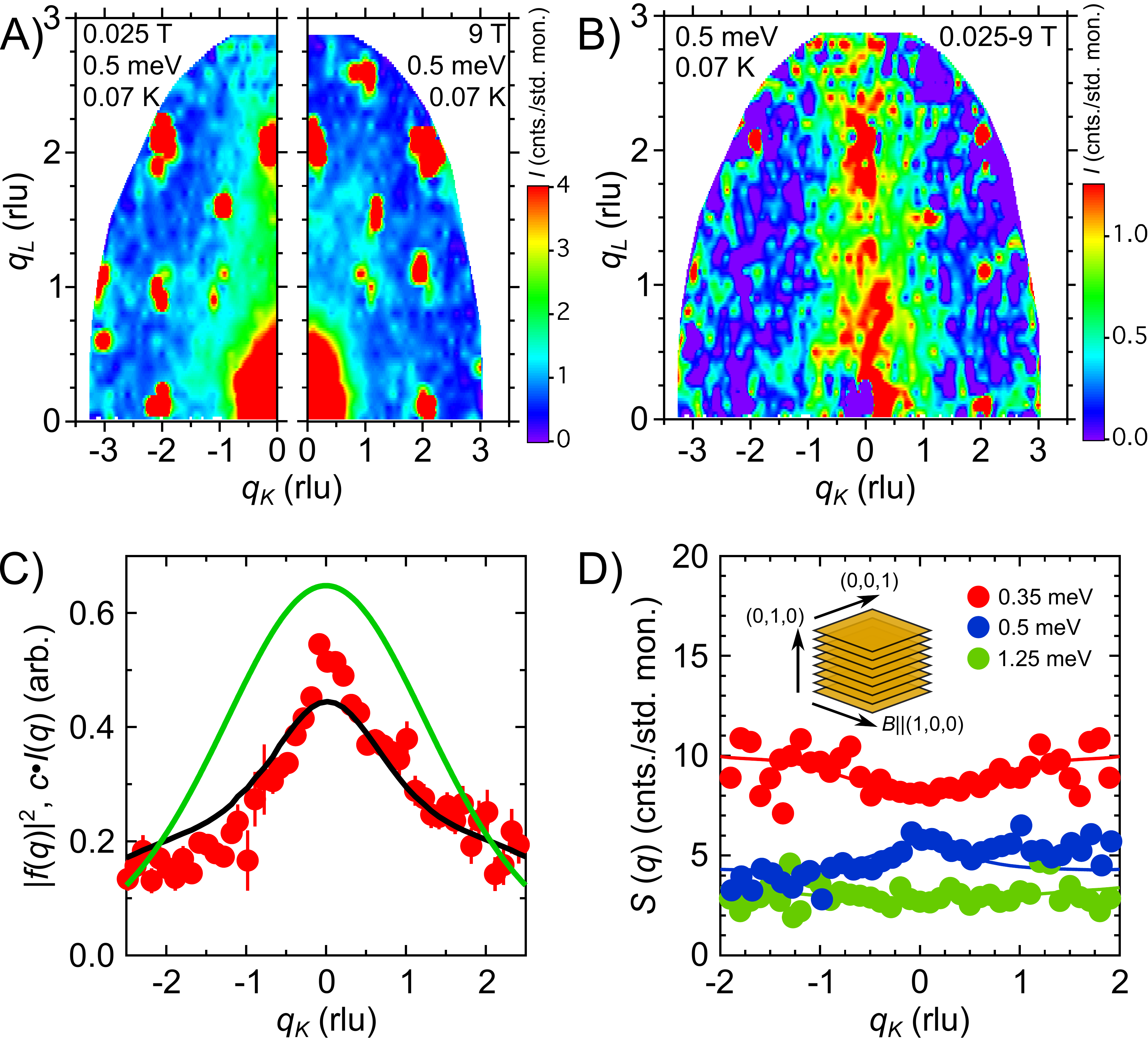}
\caption{{\bf Spatially localized magnetic fluctuations in YFe$_2$Al$_{10}$.} (A) The intensity of neutrons scattered with energy transfer $0.5~{\rm meV}$ in the [0,K,L] plane at $0.07~\rm K$ in fields of $0.025~\rm T$ (left) and $9~\rm T$ (right), and their difference $I(0~{\rm T})-I(9~{\rm T})$ (B). The tails of nuclear Bragg peaks are clearly observed in (A) at integer values of K and L. A diffuse ridge of scattering is evident along [0,0,L] at $q_{K}= 0$ reciprocal lattice units (rlu). Data are monitor normalized. (C) Wave vector $q_{K}$ dependence of the $q_{L}$ integrated intensity $I(q_{K})$ is better described by the YFe$_2$Al$_{10}$ magnetic form factor $F^{2}_{xz,yz}(q_{K})$ from electronic structure calculations (black line, also Supplementary Information) than isotropic Fe$^{2+}$ form factor (green line)~\cite{brown2006}.  Both form factors are scaled to the data.  Strong anisotropy in the intensity indicates that $d_{xz,yz}$ orbitals dominate. (D) The $T=0.07~\rm K$ structure factor $S(q_{K})$ is isolated for different fixed energies by dividing $I(q_{K})$ by  $F^{2}_{xz,yz}(q_{K})$.  Solid lines are obtained by fitting $I(q_{K})$ to a Lorentzian and dividing by the computed $F^{2}_{xz,yz}(q_{K})$, demonstrating that $S(q_{K})$ is  independent of wave vector $q_{K}$. Inset: The correspondence between the scattering wave vectors $q_{K}$ and $q_{L}$ and the $ac$-planes containing the nearly square Fe-nets in YFe$_2$Al$_{10}$. Magnetic field is oriented in the critical $ac$ plane along the (100) direction. All data were measured on MACS~\cite{Rodriguez_MST_2008}. Error bars in each figure represent one standard deviation.}
\end{figure}

The neutron intensity $I({\bf q},E)$ is the product of the magnetic form factor $F^{2}(\bf{q})$, reflecting the spatial distribution of  magnetization clouds associated with the fluctuating moments,  and the structure factor $S({\bf q},E)$, which probes correlations among moments.  The latter can be isolated (Fig. 1C,D) by comparing  $I(q_{K},E)$  to both the isotropic Fe$^{2+}$ atomic form factor~\cite{brown2006}, and to the form factor $F^{2}_{xz,yz}(q_{K})$ of the Fe $d_{xz,yz}$ Wannier orbitals, obtained from a tight binding band structure calculation (Supplementary Information).  $I(q_{K})$ falls off more quickly than the Fe$^{2+}$ atomic form factor, implying a minimal degree of Fe moment delocalization in YFe$_2$Al$_{10}$~ that is well captured by the calculations. Unlike the spherically symmetric Fe$^{2+}$ atomic form factor, $I(0,q_{K},q_{L})$ is strikingly anisotropic, and the dominance near the Fermi level of $d_{xz,yz}$ orbitals  provides a natural explanation~(Fig. S2). Once the computed form factor is removed from the measured intensity $I(q_{K},E)=F^{2}_{xz,yz}(q_{K})S(E)$, there is no further wave vector dependence of the structure factor, which is solely a function of energy $E$, $S(q,E)=S(E)$ (Fig. 1D). Since an atomic energy scale $\sim$ 1 eV controls the  the spatial distribution of the moment density in the $d_{xz}$ orbital that is reflected in the form factor, the wave vector modulation of $I(q_{K})$ is correspondingly unaffected by temperatures from $0.07 - 20~\rm K$, magnetic fields as large as $9~\rm T$, and excitation energies from $0.35 - 1.5~{\rm meV}$ (Fig. S5). Remarkably, the moments in YFe$_2$Al$_{10}$ are highly localized in space and fluctuate independently, with no sign of the spatial correlations that are a foundational element of conventional phase transitions and their $T=0$ analogs.

Despite the absence of spatial correlations among the fluctuating moments in YFe$_2$Al$_{10}$, their dynamics are manifestly QC, with the strongest scattering associated with fluctuations having the lowest energies, or longest lifetimes.  Inelastic neutron scattering experiments (Fig. 2A) reveal a gapless spectrum of excitations, where the structure factor $S(E)$, obtained from the data in Fig. 1 by integrating over $q_{K}$ (Supplementary Information), is expressed in terms of the magnetization squared $M^{2}$. The critical behavior of the energy dependence is determined by plotting the inverse of $M^{2}-C$, where $C$ is a small and energy independent contribution to the moment, as a function of $E^{1.4}$, indicating that the QC dynamics are a continuum that extends to the lowest energies probed in this experiment (Fig. 2B). Since $M^{2}$ must remain finite, the QC energy dependence $S(E)\sim E^{-1.4}$ cannot extend to $E\rightarrow0$, but must either culminate in order, absent above 0.07 K~\cite{park2011}, or be controlled by a cutoff energy that perhaps  reflects the inevitable fine tuning that would be required to bring YFe$_2$Al$_{10}$ exactly to the QCP.  By expressing $M^{2}$ in absolute units, we see that the fluctuating local moments responsible for the scattering in the energy window of our  experiment 0.35~ meV - 1.5~ meV have magnitudes of $\sim 0.3-0.4~\mu_{B}/{\rm Fe}$, similar to the local moment magnitude deduced from fitting the Curie-Weiss law to the static susceptibility  $\chi_{0}(T)$ in the temperature range $100 - 750~\rm K$~\cite{park2011}. The energy independent scattering $C$ likely reflects the presence of a broad and weakly correlated band of quasiparticle excitations, as implied by the modest Pauli susceptibility and Sommerfeld coefficient reported for YFe$_2$Al$_{10}$~\cite{park2011}. The breadth of this band is estimated as $\sim 0.7 ~{\rm eV}$(Fig.~2A), which is the energy where the integral of the fit to the experimental data reaches the square of the full spin S=2 Fe$^{2+}$ moment  $M^{2}=24~\mu_{B}^{2}$.

\begin{figure}
\centering
\includegraphics[width=.8\linewidth]{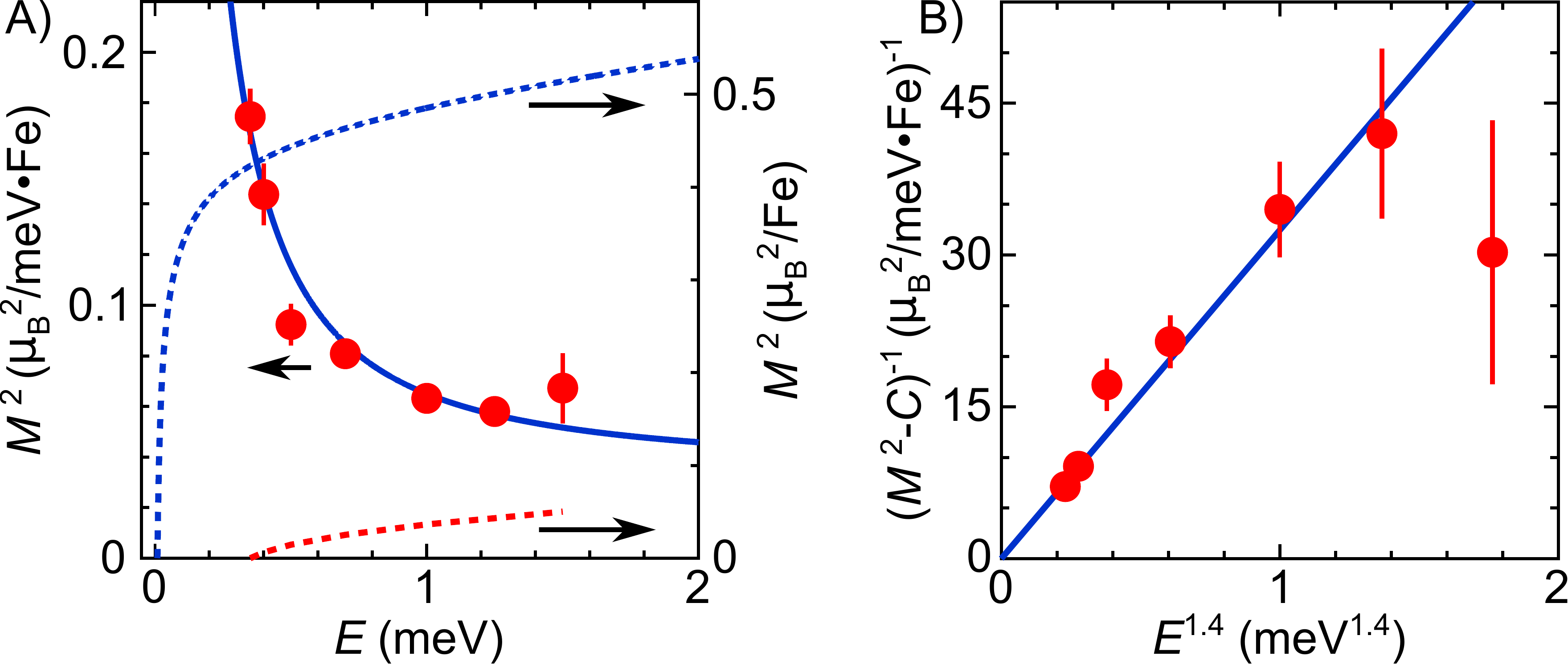}
\caption{{\bf A quantum continuum in YFe$_2$Al$_{10}$.} (A) The energy dependence of the magnetization squared $M^{2}$ of the fluctuating moments in YFe$_2$Al$_{10}$  at $0.07~\rm K$ and $B=0.025~\rm T$. For details of the normalization, see the Supplementary Information. The solid blue line is a fit to the data where $M^{2}(E)= C + aE^{-1.4}$, with $C=0.034~\mu_{B}^{2}/{\rm meV}~Fe$.  The dashed red line is the integral over the measured QC fluctuations $aE^{-1.4}$, while the dashed blue line represents the integral over the power law fit to $M^{2}(E)$ for $E>k_{B}T$. (B) The inverse of $M^{2}-C$ is plotted as a function of $E^{1.4}$. Blue line indicates the best linear fit. Error bars in both figures represent one standard deviation.}
\end{figure}

Conventionally, proximity to a phase transition results in the transfer of spectral weight to lower energies. Something very different occurs in YFe$_2$Al$_{10}$, where the $q_{L}$ integrated scattering $I(q_{K},E)$ (Fig. 3A) as well as the associated $S(E)$ (Fig. 3B) are constant over almost three decades of temperature from 0.07 K to 20 K. This simple observation has remarkable consequences.  Namely, the principle of detailed balance gives \mbox{ $S(E,T) \sim (1-exp(-E/k_{B}T))\chi''(E,T)$,} where $k_{B}$ is Boltzmann's constant. The detailed balance factor $(1-exp(-E/k_{B}T))$ is manifestly a function of $E/k_{B}T$, and thus the imaginary part of the dynamical susceptibility $\chi''(E,T)$ must also be a function of $E/k_{B}T$ that cancels the temperature dependence of the detailed balance factor. QC fluctuations having no energy scale other than temperature itself is the hallmark of QC phase transitions ~\cite{varma1989,hayden1991,keimer1991,aronson1995,schroeder2000,montfrooij2003,varma2015,inosov2010, kim2015}, and Fig. 3C shows that $\chi''=\chi''(E/T)$ in YFe$_2$Al$_{10}$~ as well.  Because our measurements in YFe$_2$Al$_{10}$~ are carried out over such a broad range of energies and temperatures, it is also possible to demonstrate that these data collapse onto a single universal curve by plotting $\chi''T^{1.4}$ as a function of $(E/k_{B}T)$ (Fig. 3D), where the universal curve is well reproduced by the expression $\chi''(E,T)T^{1.4}\propto x^{-\Delta}\tanh(x)$, where $\Delta=1.4$.  In previously investigated systems the $E/T$ scaling is always associated with the collapse of magnetic order, and it is only observed over a limited range of wave vectors that are associated with incipient magnetic order.  In contrast, the $E/T$ scaling of $\chi''$  in YFe$_2$Al$_{10}$  extends over the entire range of wave vectors accessed in this experiment, amounting to more than an entire Brillouin zone.

\begin{figure}
\centering
\includegraphics[width=.8\linewidth]{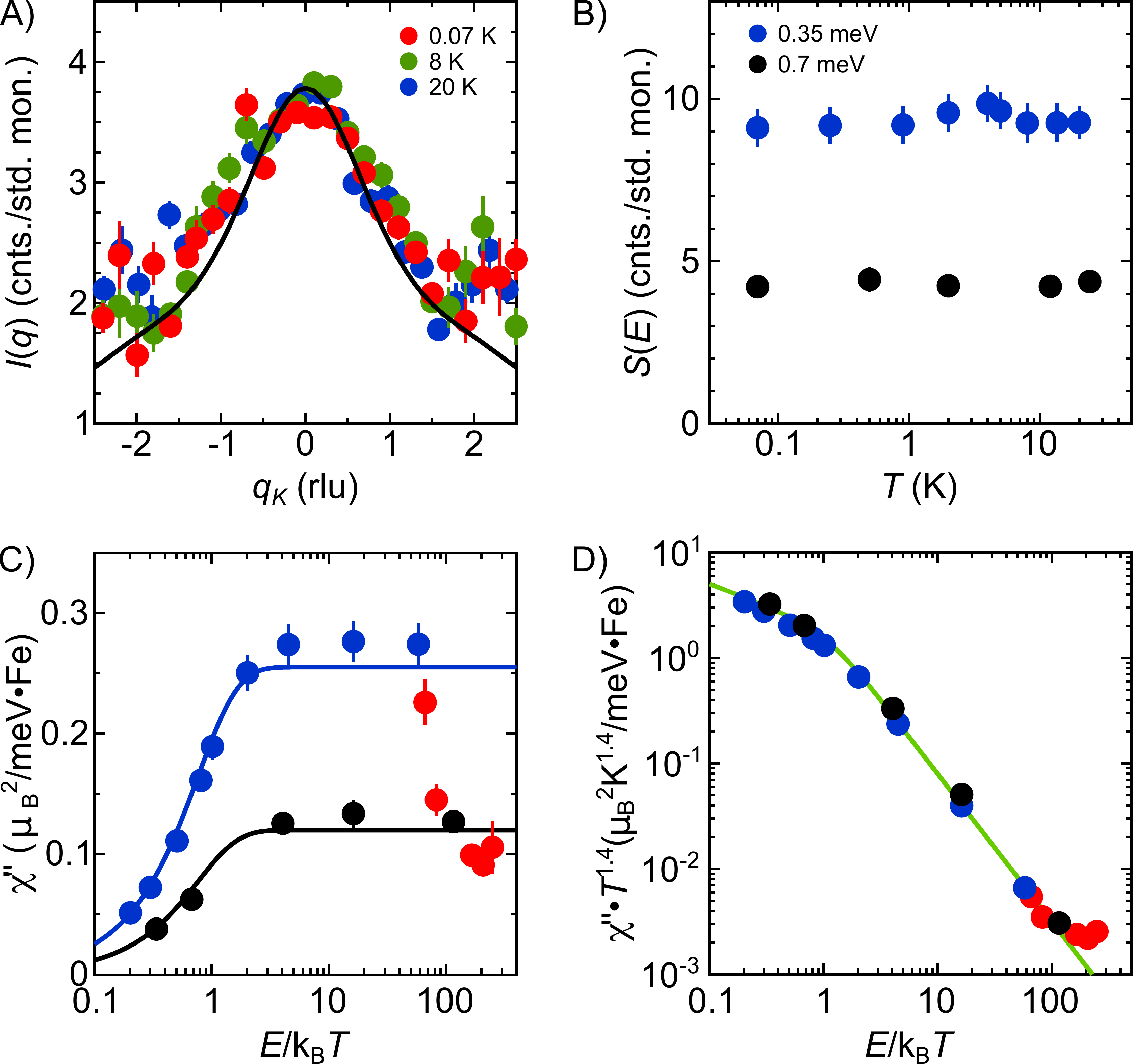}
\caption{{\bf  E/T Scaling of the magnetic dynamics in  YFe$_2$Al$_{10}$.} (A) The $q_{L}$ integrated scattering $I(q_{K})$ for an energy transfer of 0.35 meV is presented at temperatures 0.07 K (red circles), 8 K (green circles), and 20 K (blue circles). Black line is the scaled computed form factor $F^{2}_{xz,yz}(q_{K})$. (B). $S(E)$ is obtained by integrating $S(q,E)$ over experimental values of the wave vectors $q_{L}$ and $q_{K}$. Within the accuracy of our measurements, $S(E)$ is independent of temperatures in the range 0.07 K - 24 K, for the fixed energies $E=0.35~{\rm meV}$ (blue circles) and $E=0.7~{\rm meV}$ (black circles). (C) The principle of detailed balance, $\chi''(E,T) \sim S(E,T)(1-exp(-E/k_{B}T)$  is used to relate $S(E,T)$ to the imaginary part of the dynamical susceptibility $\chi''(E,T)$, which has also been integrated over these same wave vectors. $\chi''(E,T)$ is plotted at different temperatures 0.07 K - 24 K for energy transfers of 0.35 meV (blue circles) and 0.7 meV (black circles), and for a range of different energy transfers 0.35 meV - 1.5 meV at 0.07 K (red circles). The solid lines are fits to the expression $\chi'' \propto \tanh(x)$, where $x=E/k_{B}T$.  (D) The data in (C) can be collapsed onto a single universal curve when $\chi''T^{1.4}$ is plotted as a function of $E/k_{B}T$. The solid green line compares the scaled data  $\chi''T^{1.4}$ to the function $x^{-\Delta}\tanh(x)$, where $x=E/k_{B}T$, and $\Delta$=1.4. Error bars in each figure represent one standard deviation.}
\end{figure}

The energy and temperature dependencies of $\chi''$~ provide the needed connection between the neutron scattering measurements and the previously reported temperature dependence of the static susceptibility $\chi_{0}(T) \sim T^{-1.4}$ ~\cite{wu2014}, since the Kramers-Kronig relation gives
\begin{equation}
\chi_{0}(T)=\int dE ~\chi''(E,T)/E= T^{-\Delta}\int {\rm d}x~\tanh(x)/ x^{1+\Delta} 
\end{equation}
\noindent
where $\chi''E^{1.4}\propto \tanh(x)$, and $x=E/k_{B}T$. Agreement between these two independent determinations of $\chi_{0}(T)$ requires that $\Delta=1.4$, a value that is wholly within the experimental bounds of the neutron scattering experiment (Fig. 2B). In addition, the integral itself must remain finite. The strong divergence of $\chi''(E)$, implies that it cannot extend to arbitrarily low energies and temperatures, and a  proposal for a particular energy and temperature cutoff is compared to the scaled data in the Supplementary Information.  For the range of temperatures and energies accessed in our experiments, the matching energy and temperature dependencies of the neutron scattering and magnetic susceptibility measurements imply that both measurements probe the same QC fluctuations. It is worth pointing out that the strong energy and temperature divergencies of $\chi_{0}$ and $\chi''$ are inconsistent with an appreciable role for disorder in the QC behavior of YFe$_2$Al$_{10}$~\cite{fisher1992,thill1995,castroneto2000,vojta2005,miranda2005}.

\begin{figure}
\centering
\includegraphics[width=.8\linewidth]{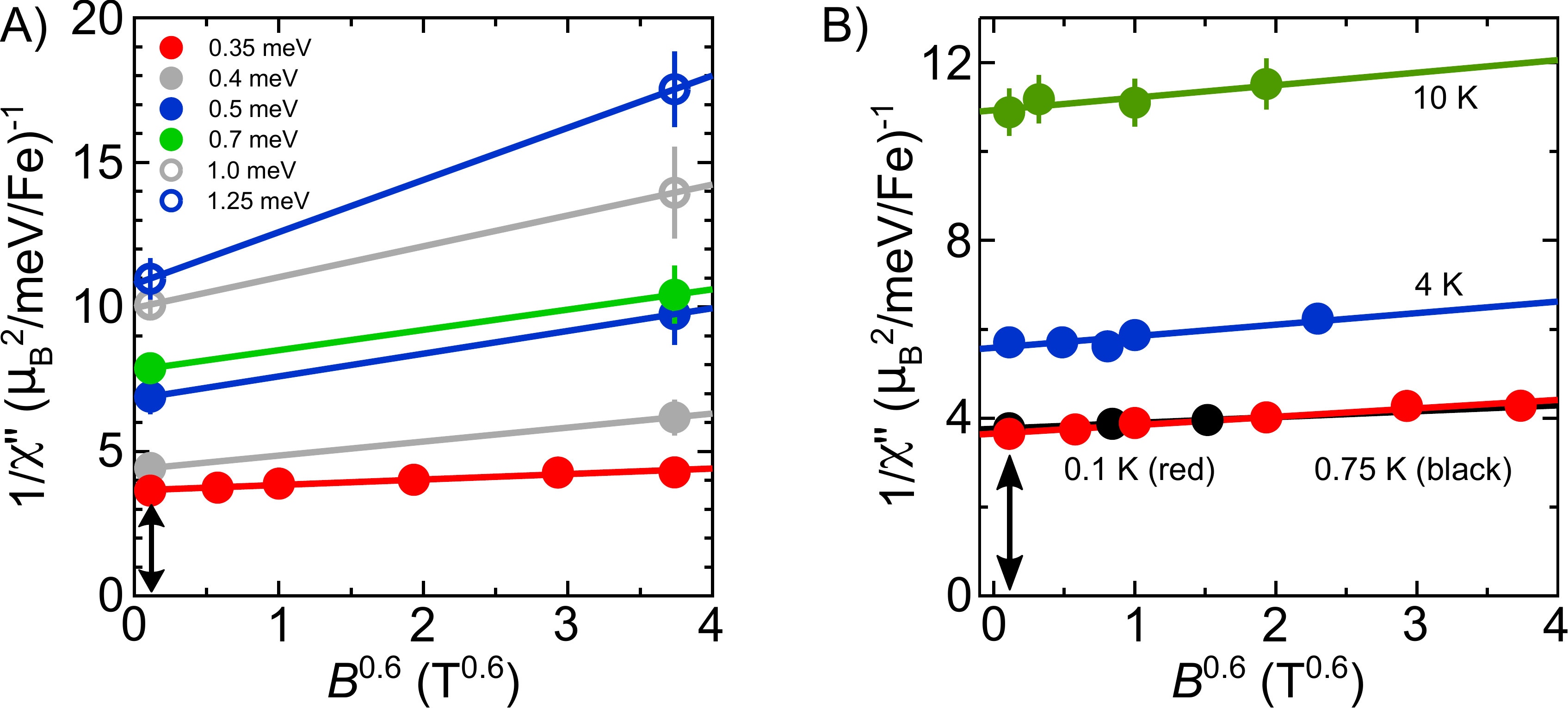}
\caption{{\bf Magnetic fields and the Quantum Critical Scaling of the dynamics in YFe$_2$Al$_{10}$.} (A). The inverse of $\chi''$~  is plotted as a function of $B^{0.6}$ to demonstrate that the expected divergence of $\chi''$~ for $B\rightarrow~0$ is cutoff by the nonzero energies 0.35 meV - 1.25 meV  of the T=0.07 K neutron scattering measurements. (B)~$\chi''$~ measured with $E=0.35~{\rm meV}$ is plotted as a function of $B^{0.6}$ for different temperatures, as indicated. Double ended arrows in A,B indicate the same values of $1/\chi"$, measured at $T=0.07~\rm K$ and $E=0.35~{\rm meV}$ (Fig.~2A), which provides the cutoff for the data in both (A) and (B). Error bars in each figure represent one standard deviation.}
\end{figure}

The structure, symmetries, and interactions present in a given material determine the conditions under which a $T=0$ phase transition may occur, and so  modifications to these quantities via pressure, stress, or composition will affect the tendency to order. Increasing temperature weakens QC fluctuations, as would magnetic fields if the QCP corresponds to the onset of magnetism.  Scaling analyses of the static susceptibility \mbox{$\chi_{0}\sim T^{-1.4}f(T/B^{0.6})$ }have shown that a single variable $T/B^{0.6}$ controls the QC fluctuations for a wide range of fields and temperatures in YFe$_2$Al$_{10}$ \cite{wu2014}. This variable is also observed in neutron scattering measurements. $\chi''$ is a function of $(E,T,B)$, whose properties can in general only be determined in particular limits where there is a dominant scale. These neutron scattering experiments directly probe the response to the fluctuating fields associated with the QCP in YFe$_2$Al$_{10}$ for energies that are, for the most part, larger than the thermal energy $k_{B}T$ and the Zeeman energy $g\mu_{B}B$.  The simplest case (Fig.~2B) takes $B=0$ and $T=0.07~\rm K$, where the energy dependence dominates and gives $1/\left(\chi''-\tilde{C}\right) = \tilde{a}E^{1.4}$, with $\tilde{C}=0.053~\mu_{B}^2/{\rm meV~Fe}$.  At $T=0.07~\rm K$,  the thermal energy is irrelevant and it is possible to probe the field dependence of $S$, when the condition $g\mu_{B}B\leq E$, is for the most part realized, assuming $g=2$. Here, the same energy dependence provides a cutoff in the divergence of the field dependence of $\chi''$, i.e. $1/\chi''=C_0+bB^{0.6}$ (Fig. 4A,B), where $C_0$ represents the scattering at $E=0.35~{\rm meV}$ (Fig. 2) articulating the field suppression of the scattering first demonstrated in Fig. 1A. Our data with $E=~0.35~{\rm meV}$ and temperatures having $k_{B}T$ both larger and smaller than $E$ finds evidence for the same energy dependent cutoff, as well as a separate cutoff that increases with temperature $T$. The $T/B^{0.6}$ scaling found in $\chi_{0}(T)$ is observed when the excitation energy E is small compared to the thermal and magnetic field scales, a regime that is largely unaddressed in our neutron scattering measurements. These measurements indicate that the QC dynamics extend over a considerable range of fields and temperatures, but they appear to be controlled by proximity to a very specific QCP at $T=0,B=0$, arguing against an extended non-FL phase.

The previously reported scaling analysis made it clear that YFe$_2$Al$_{10}$  is naturally located by its composition to be very close to a $T=0$ phase transition. The neutron scattering measurements reported here reveal that this phase transition is highly unconventional. Specifically, the near divergence of $S(E)$ as $E\rightarrow0$ shows that QC magnetic fluctuations with a time scale $\xi_{\tau}$ dominate at the low energies probed in these experiments, while spatial correlations $\xi_{r}$ among these moments are absent. This violates the foundational property of conventional phase transitions~\cite{hohenberg1977}, where $\xi_{r}$ and $\xi_{\tau}$ are related by the dynamic exponent z, $\xi_{r}^{z}=\xi_{\tau}$ .  An intriguing alternative has recently been suggested, where a topological phase transition could produce anomalously weak spatial correlations, as well as reproducing several of the experimental findings in YFe$_2$Al$_{10}$~\cite{hou2016,zhu2016}.

Our major finding is that the excitations detected by our neutron scattering measurements in YFe$_2$Al$_{10}$  are those of individual and highly localized magnetic moments, each fluctuating independently with the same anomalous spectrum, without any evidence for a nearby broken symmetry. The low temperature divergencies of quantities like the magnetic susceptibility, specific heat, and the electrical resistivity all attest to the breakdown of normal metallic behavior, which we now know occurs in the absence of magnetic order at temperatures as low as 0.07 K. The observation of $E/T$ scaling in the neutron scattering measurements indicates that the magnetic excitations are fundamentally modified relative to the damped spin waves or the continuum of single particle excitations that are expected near a classical magnetic phase transition.

The small but localized moments identified by both the Curie-Weiss susceptibility and the neutron scattering measurements imply that YFe$_2$Al$_{10}$  forms very close to an electronic localization transition. As was demonstrated in both Fe and Mn pnictides and  chalcogenides~\cite{haule2009,yin2011,mcnally2014}, such moments result from Hunds and Coulomb interactions that provide electronic correlations that are potentially strong enough to localize one or more Fe d-orbitals in YFe$_2$Al$_{10}$. Since the localized moments emerge from a relatively flat band (Fig. S2), it is likely that the form factor of the moments, which encodes the orbital content,  will dominate the {\bf q}-dependence of the scattering, just as we have observed. The stabilization of the Fe moments is envisaged as a continuous crossover or transition between a coherent metallic state where the localized moments are wholly quenched, and a state where this compensation has failed, leading to incoherent and localized magnetic moments~\cite{haule2009}.  A Mott-like transition could ensue at $T=0$ for a critical interaction strength, accompanied by QC fluctuations between these two topologically distinct states that are degenerate at the QCP.  The comparison of the measured and computed form factors suggests that it is the $d_{xz,yz}$  orbitals that are most localized in YFe$_2$Al$_{10}$, while the other orbitals are represented as delocalized and weakly correlated electronic states that result in the overall metallic character of YFe$_2$Al$_{10}$, evident from the temperature dependence of the electrical resistivity, as well as the modest Pauli susceptibility and Sommerfeld constant. Consequently, it  seems possible that YFe$_2$Al$_{10}$  is very close to an Orbital Selective Mott Transition (OSMT)~\cite{anisimov2002,demedici2005,demedici2009,mvojta2010}, and that it is QC fluctuations between these phases at $T=0$ that lead to the non-FL properties of YFe$_2$Al$_{10}$. Detailed investigations of the Fermi surface in YFe$_2$Al$_{10}$, ideally as pressure or another nonthermal parameter tunes the localized moments to extinction, will be required to further evaluate this proposal.

For now, the nature of the $T=0$ phase transition that drives the quantum critical behavior that is so dominant in YFe$_2$Al$_{10}$  remains unknown, although its consequences are transformative. Neutron scattering provides a powerful and direct means to show that this phase transition is  not of the conventional Landau-Ginsburg-Wilson type. Unlike previously studied systems where similar measurements found that QC behavior was never wholly free of the magnetic correlations associated with proximate magnetic order, the complete absence of these correlations in YFe$_2$Al$_{10}$  indicates that here the QCP stands alone, and is definitively of a type that has never been observed before.  YFe$_2$Al$_{10}$  is almost unique, in that no fine tuning is required to access its QCP, which affects a remarkably broad range of temperatures and fields. In this sense, it might be considered the d-electron analog of $\beta-{\rm YbAlB}_{4}$~\cite{matsumoto2011,tomita2015}. Since no bulky pressure apparatus and no potentially disruptive disorder from compositional variation are necessary in YFe$_2$Al$_{10}$  to fine tune the QCP, our results open the door to further explorations of the nature and properties of this most novel quantum phase transition, using the most powerful spectroscopic and imaging tools at our disposal.

\section*{Methods}
\subsection*{Samples and Experimental Setup}To measure the excitation spectrum of YFe$_2$Al$_{10}$, neutron scattering measurements were carried out in the $0,q_{K},q_{L}$ scattering plane on the Multi Axis Crystal Spectrometer (MACS) instrument at the Center for Neutron Research at the National Institute of Standards and Technology~\cite{Rodriguez_MST_2008}.  The sample consisted of two co-aligned single crystals of YFe$_2$Al$_{10}$  with a total mass of $2 ~\rm g$,  mounted in a dilution refrigerator equipped with a superconducting magnet with an 11 T vertical field aligned with the (100) crystal direction. In order to reduce background scattering in the double focusing mode, we used a  3.3 cm $\times$ 7.7 cm beam mask to focus the neutron beam on the 1.5 cm $\times$ 2.5 cm sample.  For all measurements, a small bias field of about $0.025~\rm T$ was used, suppressing superconductivity of the aluminum sample holder at low temperatures and for consistency at temperatures above $T_c$ of aluminum, 1.2 K.  Undesired background scattering was eliminated by setting the dark angle of the magnet at 90 degrees away from the (010) direction and using a final neutron energy $E_f$ = 3.0 and 3.7 meV ($\lambda_f = 5.22$ and 4.70 \AA, respectively).  Be filters were used between the neutron source and the sample, while Be (for $E_f=3.0~{\rm meV}$) or BeO (for $E_f=3.7~{\rm meV}$) filters were used between the sample and detector.  All reciprocal lattice vectors are indexed as ($q_H$ $q_K$ $q_L$), with  reciprocal lattice units $q_K=2\pi /b = 0.62$ \AA$^{-1}$ and $q_L=2\pi /c = 0.70$ \AA$^{-1}$.

\subsection*{Data Analysis}The quantities of interest determined in our neutron scattering measurements at a given wavector $\bf q$ and energy $E$ are the magnetic structure factor $S\left(\textbf{q}, E\right)$ and the imaginary part of the magnetic susceptibility $\chi''\left(\textbf{q}, E\right)$.  Both relate to our measured neutron scattering intensity $I\left(\textbf{q}, E\right)$ in a straightforward way~\cite{Squires_book_2012}.  $S\left(\textbf{q}, E\right)$ is determined by dividing $I\left(\textbf{q}, E\right)$ by the square of the magnetic form factor $F^2(\textbf{q})$,
$S\left(\textbf{q}, E\right)\sim I\left(\textbf{q}, E\right)/F^2(\textbf{q})$, while $\chi''\left(\textbf{q}, E\right)$ is related to $S\left(\textbf{q}, E\right)$ by the principle of detailed balance, where $\chi''\left(\textbf{q}, E\right) \sim S\left(\textbf{q}, E\right) \left(1-e^{-E/k_BT}\right)$ for a given temperature $T$.

Our measurements on the Multi Axis Crystal Spectrometer (MACS) instrument at the Center for Neutron Research at the National Institute of Standards and Technology~\cite{Rodriguez_MST_2008} (see Fig.~1A of the main text for an example of the data) acquire the $\bf q$ dependence of the scattered neutron intensity  at different fixed energy transfers, temperatures, and magnetic fields in the ${\bf q}=$[0, K, L] plane (the $q_K$--$q_L$ plane).  After normalizing the measured intensity by the incident neutron flux, areas of $\bf q$-space contaminated by tails from Bragg reflections were masked.  The results were integrated (i.e. numerically summed) along the $q_L$ direction over the range L=[0.8, 1.8] rlu, and were then normalized by that range of $q_L$, $\Delta q_L$, which covers one full Brillouin Zone, Eqn. S1. This zone was selected to minimize possible contamination from the direct beam.  This procedure yields the $q_K$ dependence of our measured intensity, $I\left(q_K, E\right)$,
\begin{equation}\label{qLInt}
I\left(q_K, E\right)=\int_{0.8~{\rm rlu}} ^{1.8~{\rm rlu}} I\left(q_K, q_L, E\right)\mathrm{d}q_L/\Delta q_L
\end{equation}

Examples of $I(q_{K},E)$ are given in Figs. 1(C), 3(A), and 4(A) of the main text and in Fig. S5 of this supplement.  $I(E)$ can be obtained with a similar integration of the $q_K$ dependence.  $S(q,E)$ and $S(E)$ are obtained from these quantities after accounting for the form factor, which is described in supplemental information.

\section*{Acknowledgements}
Part of this research was conducted at Brookhaven National Laboratory, where  W. J. G, L. S. W., and M. C. A  were supported under the auspices of the US Department of Energy, Office of Basic Energy Sciences, under contract DE-AC02-98CH1886.  I. A. Z, A. M. T, and W. X. have been separately supported under the auspices of the US Department of Energy, Office of Basic Energy Sciences, under contract DE-SC00112704.  W. X. was supported by Center for Computational Design of Functional Strongly Correlated Materials and Theoretical Spectroscopy under DOE grant DE-FOA-0001276.  Part of this work was performed at the Aspen Center for Physics, which is supported by National Science Foundation grant PHY-1066293. Access to MACS was provided by the Center for High Resolution Neutron Scattering, a partnership between the National Institute of Standards and Technology and the National Science Foundation under Agreement No. DMR-1508249.  The authors acknowledge useful discussions with C. Varma, E. Abrahams, D. MacLaughlin, L. Shu, S. Chakravarty, G. Aeppli, S. Raymond, and C. Broholm.

\section*{Contributions}
W. J. G., L. S. W, and M. C. A designed the research. L. S. W. synthesized and characterized the crystals and assembled the multicrystal sample for neutron scattering experiments. W. J. G., L. S. W., and I. A. Z. performed the experiments with the assistance of J. R.-R. and Y. Q.  W. J. G. analyzed the data. W. X. and A. M. T. provided electronic structure calculations and the form factor calculations. W. J. G. and M. C. A. wrote the paper with contributions from all authors.  The authors declare that they have no conflicts of interest.

\pagebreak

\renewcommand{\thefigure}{S\arabic{figure}}
\renewcommand{\theequation}{S\arabic{equation}}

\setcounter{figure}{0}
\setcounter{equation}{0}

\section*{Supporting Information for A Local Quantum Phase Transition  in $\mathrm{YFe}_{2}\mathrm{Al}_{10}$}

\subsection*{Normalization to Absolute Units}\

In order to compare our results to the bulk properties of YFe$_{2}$Al$_{10}$, it is necessary to estimate the magnitude of the scattering in absolute units.  The method we use~\cite{Xu_RSI_2013} was specifically developed for triple axis neutron scattering measurements, and so it is appropriate for MACS, which is best considered to be 20 coupled triple axis detectors with simultaneous diffraction detectors in each of the 20 channels~\cite{Rodriguez_MST_2008}.

The structure factor $S\left(\textbf{q}, E\right)$ is given in units of meV$^{-1}$ by

\begin{equation}\label{SinAbs}
S\left(\textbf{q}, E\right)=\left(\frac{2}{\gamma r_0}\right)^2 \frac{I\left(\textbf{q}, E\right)}{g^2F^2\left(\textbf{q}\right)e^{-2W}Nk_fR_0},
\end{equation}

where $\gamma=1.91$ is the gyromagnetic ratio of the neutron in nuclear magnetons, $r_0=2.818 \times 10^{-13}$ cm is the classical radius of the electron, g is the g-factor for the magnetic moments, $e^{-2W}$ is the Debye-Waller factor, $N$ is the number of unit cells in the sample, $k_f$ is the wave vector of the scattered neutrons, and $R_0$ is the resolution volume.  The term $\left(2/\gamma r_0\right)^2$ reduces to 13.77 b$^{-1}$, where 1 b=10$^{-24}$ cm$^2$.

Determining the resolution volume $R_0$ is not straightforward on an interconnected instrument such as MACS.  Away from any magnetic or structural scattering, the measured intensity will be given solely by the incoherent scattering in the energy window over which the measurement is made.  Elastic incoherent scattering has a very simple cross section, given by~\cite{Squires_book_2012, Xu_RSI_2013}:
\begin{equation}
I\left(\textbf{q}, E=0\right)^{inc} = \frac{N}{4\pi}\sum_j \sigma^{inc}_{j}e^{-2W}
\end{equation}
where $j$ denote all atoms in the unit cell and $\sigma^{inc}_j$ is the incoherent scattering cross section of the $j$th atom.  The energy integrated intensity measured in an experiment that scanned through $E=0$ at a position in reciprocal space with purely incoherent scattering will only include additional terms for the spectrometer resolution and incident wave vector $k_i$ and is given by:
\begin{equation}
\int I\left(\textbf{q}, E\right)^{inc}dE = \frac{N}{4\pi}\sum_j \sigma^{inc}_j e^{-2W_j} k_i R_0.
\end{equation}
One then finds
\begin{equation}
NR_0=\frac{4\pi}{k_i} \frac{\int I\left(\textbf{q}, E\right)^{inc}dE}{\sum_j \sigma^{inc}_j e^{-2W_j}}.
\end{equation}
This value of $NR_0$ can then be substituted into Eqn. S2, giving
\begin{equation}
S\left(\textbf{q}, E\right)=\frac{13.77\cdot I\left(\textbf{q}, E\right)}{g^2F^2\left(\textbf{q}\right)e^{-2W}4\pi \frac{k_f}{k_i}\frac{\int I\left(\textbf{q}, E\right)^{inc}dE}{\sum_{j} \sigma^{inc}_j e^{-2W_j}}}
\end{equation}
At low temperatures, $W\rightarrow 0$ and we take $e^{-2W}$ and all $e^{-2W_j}$ to be 1.  The incoherent scattering cross sections of the $j$ atoms in the unit cell are tabulated~\cite{NN}, with $\sigma^{inc}_{j} = $ 0.15 b for Y, 0.4 b for Fe, and 0.0082 b for Al.  The unit cell of YFe$_{2}$Al$_{10}$\ contains 4 Y atoms, 8 Fe atoms, and 40 Al atoms, giving $\sum_{j}\sigma^{inc}_j=4.128$ b for one unit cell.  The form factor $F(\textbf{q})$ is accounted for in the method described in the next section of this supplement, $I(\textbf{q}, E)$ is our experimentally measured inelastic intensity, while $k_{i,f}$ are the initial and final (inelastic) neutron wave vectors used in these experiments.

What remains to be determined is the integral  $\int I\left(\textbf{q}, E\right)^{inc}\mathrm{d}E$.  The strategy we employ for estimating this quantity is to use the separate diffraction detectors on the MACS detector bank~\cite{Rodriguez_MST_2008}.  These detectors monitor the elastic scattering channel while the inelastic scattering measurements are conducted, so they provide a measurement at $E=0$ with the same incident energy (and thus $k_i$) and over the same incident  neutron flux as he inelastic measurements.  The elastic, incoherent scattering was measured at $\bf{q}=$[0, 1.5, 0], far away from any nuclear Bragg reflections or the scattering of experimental interest.  Although we do not have an energy scan through $E=0$ to determine the energy integrated intensity as MACS measures wavevector dependence at a constant $E$, we estimate it by assuming a Gaussian line shape with a half-width of 10\% of the incident neutron energy, a resolution typical of a triple-axis instrument.  The integral of this gaussian estimates $\int I\left(\textbf{q}, E\right)^{inc}{\rm d}E$ , completing the determination of  $S\left(\textbf{q}, E\right)$.

This method gives $S\left(\textbf{q}, E\right)$ in units of meV$^{-1}$, which can subsequently  by related to the square of the magnetic moment per Fe atom by
\begin{equation}
M^2\left(\textbf{q}, E\right) = g^2\mu_B^2S\left(\textbf{q}, E\right)
\end{equation}
in units of $\mu_B^2/\mathrm{meV}\! \cdot \! \mathrm{Fe}\: \mathrm{atom}$ .  The imaginary part of the magnetic susceptibility is related to $M^2\left(\textbf{q}, E\right)$ through the fluctuation dissipation theorem:
\begin{equation}
\chi''\left(\textbf{q}, E\right) = \frac{\pi}{2}M^2\left(\textbf{q}, E\right)\left(1-e^{-E/k_BT}\right).
\end{equation}

It is important to note that this procedure was applied equivalently to all data and any uncertainty in this procedure, which is derived mainly from the estimate of$\int I\left(\textbf{q}, E\right)^{inc}{\rm d}E$, is present as a systematic uncertainty in all data sets and shifts the absolute scale equivalently for all points presented here.  Scattering from the sample environment contributing inordinately to our incoherent measurement will lead to an underestimate of the absolute magnitude for example, while the uncertainty in the absolute magnitude of the incoherent scattering, incoherent scattering width, or g-factor (which we take to be 2) could shift our estimate either way.  In any case, this systematic error does not change the fundamental conclusions reached in the main text of this report.

\subsection*{First principles calculations of the electronic structure in YFe$_{2}$Al$_{10}$}
The first principles calculations are performed using the full electron scheme of the density functional theory (DFT) with the generalized gradient approximation as implemented in WIEN2k~\cite{blaha2001}. The Brillioun zone and the high-symmetry points used in the DFT calculations are shown in Fig. S1. For convenience, the Fe $3d$ orbitals are written in Cartesian coordinates with $x \parallel \mathbf{a}$, $y \parallel \mathbf{c}$, and $z \parallel \mathbf{b}$, where $\mathbf{a}$, $\mathbf{b}$, and $\mathbf{c}$ are the crystallographic axes following Figure 1A in Ref.~{\cite{wu2014}}. With this convention, $d_{xy}$ and $d_{x^2-y^2}$ lie in the $x-y$ plane, $d_{xz}$ lies in the $xz$ plane, $d_{yz}$ lies in the $yz$ plane, and $d_z^2$ aligns with $z$ axis.

\begin{figure}
\centering
\includegraphics[width=.8\linewidth]{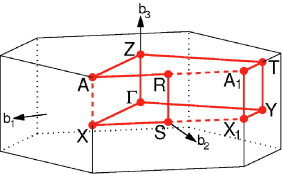}
\caption{{\bf The Brillouin zone} The Brillioun zone and high-symmetry points in YFe$_2$Al$_{10}$~\cite{setyawan}.}
\label{fig:fig1}
\end{figure}

The band structures with Fe $3d$-character are shown in Fig. S2. The band structure reflects the strong hybridization between the Fe $3d$ orbitals and bands of neighboring Al atoms. The orbital character of the low energy bands is dominated by $d_{xz}$ and $d_{yz}$. In particular, the $d_{yz}$ orbital forms a narrow band along $\Gamma-X-S-X_1-Y$ slightly above the Fermi level, and a flat band slightly below near $S$. The narrow band with $d_{xz}$ character spans $R-S-\Gamma$.

\begin{figure}
\centering
\includegraphics[width=.8\linewidth]{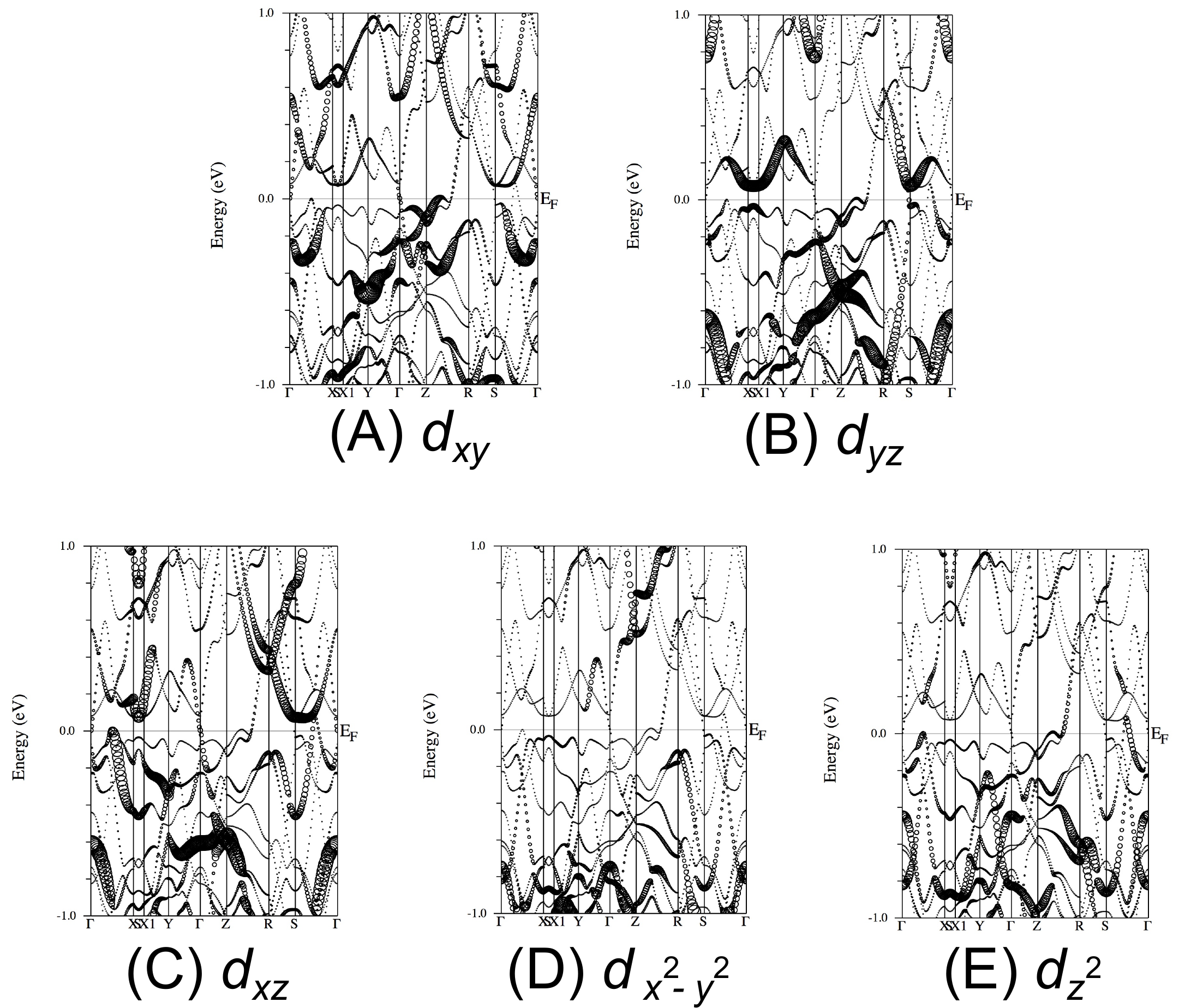}
\caption{{\bf Electronic band structure with Fe $3d$-character.}  The band structure with Fe $3d$-character is shown with the different orbitals being:  (A) $d_{xy}$,  (B) $d_{yz}$, (C) $d_{xz}$, (D) $d_{x^2-y^2}$, and (E) $d_{z^2}$.}
\label{fig:fig2}
\end{figure}

We compute the maximally localized Wannier functions, using the wannier90 code~\cite{PRB56,PRB65,wannier} and the Wien2Wannier interface~\cite{kunevs2010}. The band structure of the tight-binding model derived from the Wannier functions captures the band structure near the Fermi level from density functional calculations rather well. For a given Wannier function $\psi_{\mathbf{R}}(\mathbf{r})$ centered at $\mathbf{R}$, the form factor $F(\bf q)$ is calculated by
\begin{equation}
  F(\bf q) = \int d^3\mathbf{r} \exp\left( i\bf Q \cdot \mathbf{r} \right) \rho_{\mathbf{R}}(\mathbf{r}),
\end{equation}
where $\rho_{\mathbf{R}}(\mathbf{r}) = |\psi_{\mathbf{R}}(\mathbf{r})|^2$ is the electron density of the Wannier function. Fig. S3 shows the $F^2(\bf q)$ in the [0KL] plane for the Fe $3d$ orbitals. For each $3d$ orbital, $F^2(\bf q)$ is the average over that of all Fe atoms in the orthorhombic unit cell. For comparison, we also show the isotropic $F^{2}(\bf q)$ expected for an isolated Fe$^{2+}$ ion (Fig. S3(A)).

\begin{figure}
\centering
\includegraphics[width=.8\linewidth]{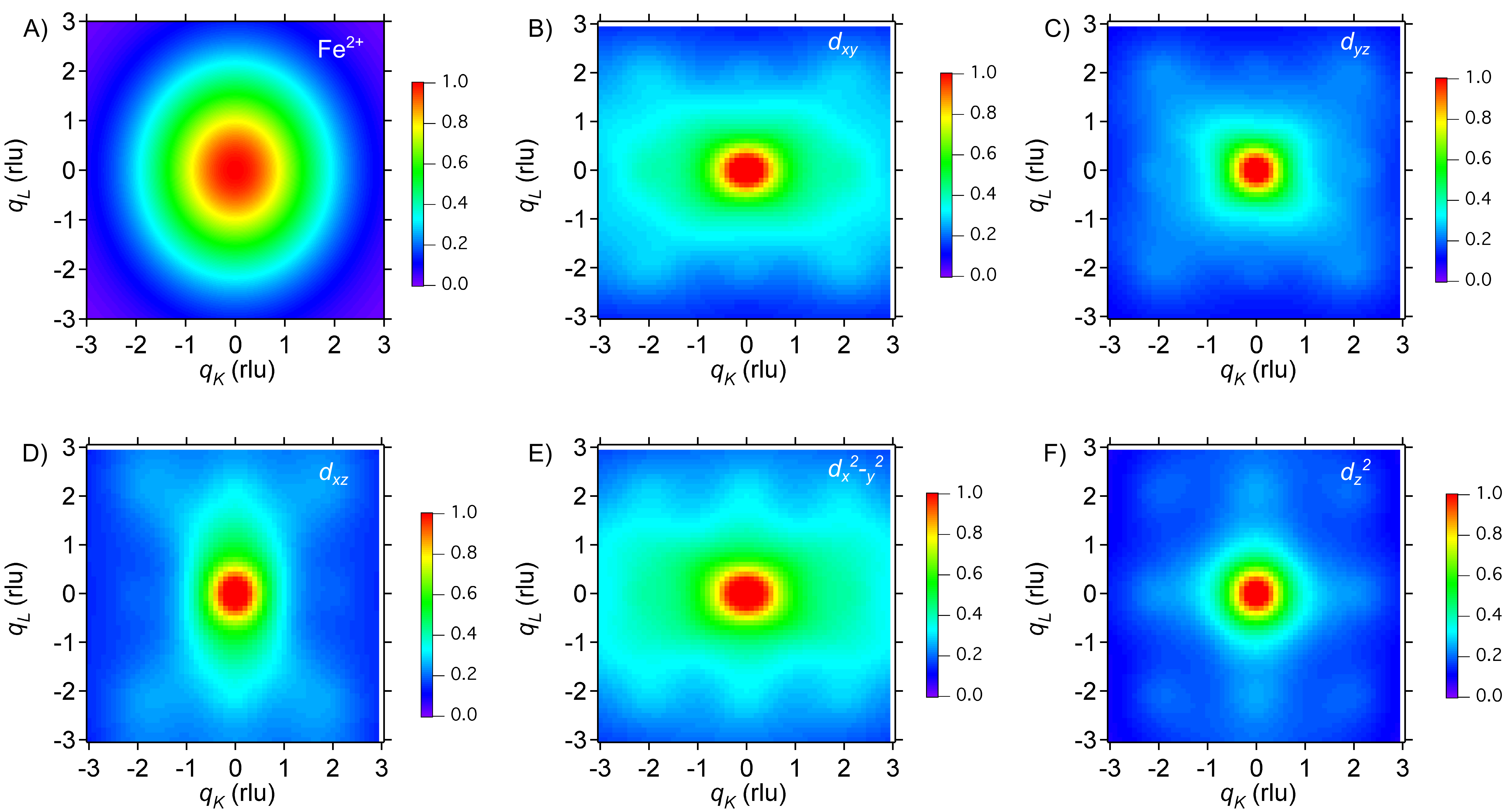}
\caption{{\bf Iron form factors in the [0KL] plane} (A) The square of the form factor $F^2(\bf q)$ for Fe$^{2+}$ calculated as described in \cite{brown2006}.  (B-F) Averaged $F^2(\bf q)$ in the [0KL] plane from Wannier functions for Fe $3d$ basis for the (B) $d_{xy}$ orbital, (C) $d_{yz}$ orbital, (D) $d_{xz}$ orbital, (E) $d_{d^2-y^2}$ orbital, (F) and the $d_{z^2}$ orbital.}
\label{fig:fig3}
\end{figure}

By comparing the form factors presented in Fig. S3 with the data measured in the [0KL] plane (Fig. 1A, main text) it is clear that the data most resemble the form factor associated with the $d_{xz}$ orbital, and given the near tetragonal symmetry of YFe$_{2}$Al$_{10}$, measurements in
the [HK0] direction would presumably resemble the form factor of the $d_{yz}$ orbital. The anisotropy of the form factor in YFe$_{2}$Al$_{10}$~ indicates that the $d_{xz,yz}$ orbitals are the primary origin of the Fe magnetism in YFe$_{2}$Al$_{10}$. What is more, the measured $q_K$ dependence of the scattered
intensity (Fig. 1C main text) is very well described by the computed form factor, indicating that the most correlated states near the Fermi level have substantial $d_{xz,yz}$ character. The computed and measured form factors have similarly strong {\bf q}-dependencies as the Fe$^{2+}$ form factor, indicating that they all are quite localized. The $q_{K}$ dependence of the form factor that we use in our data analysis is obtained from the calculation of $F^{2}(q)$ for the $d_{xz}$ orbital by integrating in a fashion similar to that of Eqn. S1, but with the calculated quantity in place of the measured intensity.

\subsection*{Comparison of Data to Form Factor for Wavevectors along $q_L$}
The comparisons of the raw measured data to the calculated form factor described in the main text of this paper focus on the $q_K$ dependence, the direction in reciprocal space with the greatest modulation in the raw data.  The comparison of our measurements to the form factor also works well for the perpendicular $q_L$ dependence.  In Fig. S4(A, B) we show the raw measurement of the $q_L$ dependence summed over a narrow strip of $q_K$ around $q_k=0.5$ rlu at both $E=0.5$ and 1.0 meV and $B=0$ and 9 T.  There is a slow decrease in the size of the signal with increasing $q_L$, qualitatively consistent with the calculated form factor for the $d_{xz}$ orbital.  When these cuts are divided by $F^2\left(q_L\right)$, the results are $\textbf{q}$ independent within the scatter of the data, indicating that the momentum dependence in this direction is completely described by the $d_{xz}$ orbitals originating with the Fe atoms in YFe$_{2}$Al$_{10}$\ (Fig. S3(D)).

\begin{figure}
\centering
\includegraphics[width=.8\linewidth]{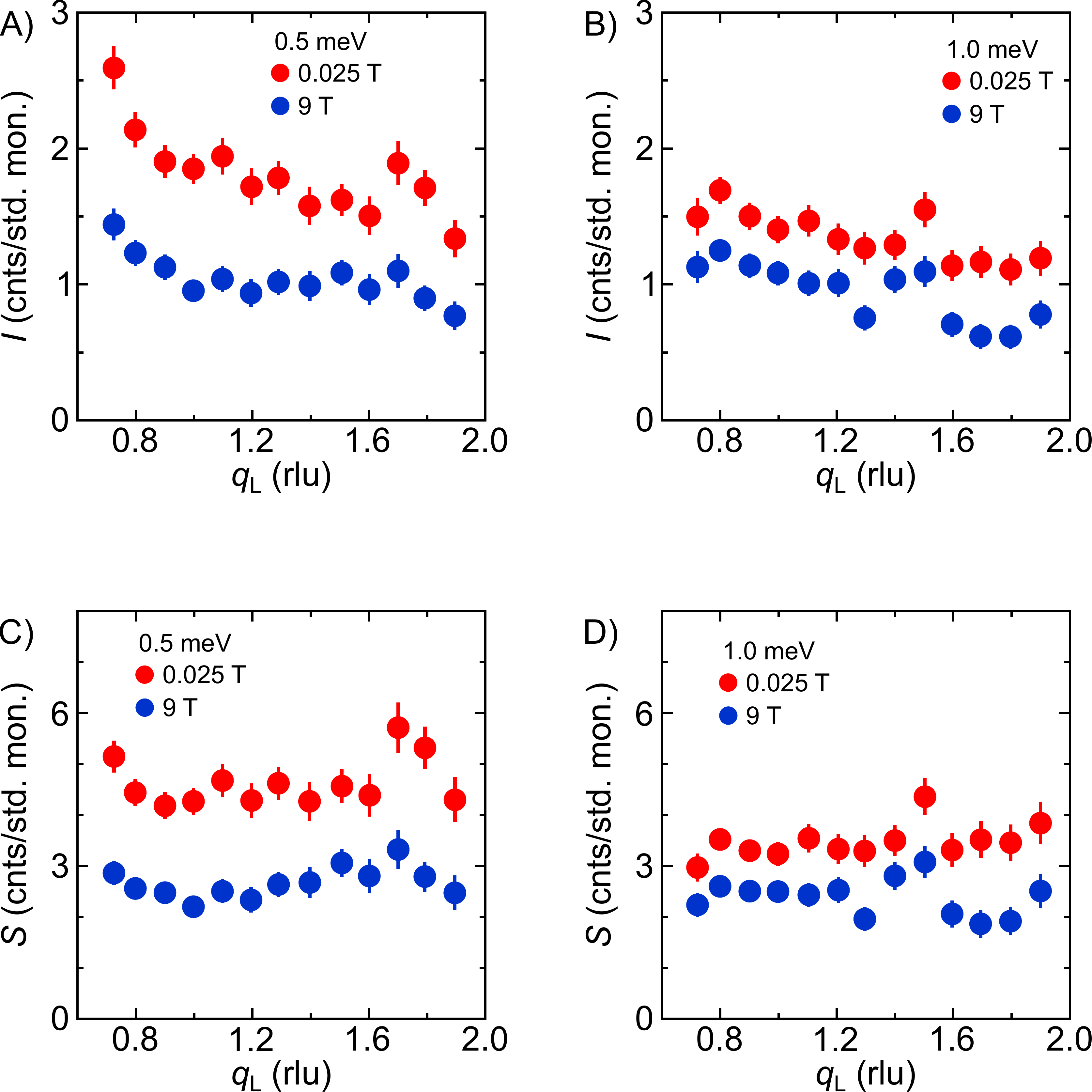}
\caption{{\bf $q_L$ dependence of scattering.} (A,B) The $q_L$ dependence of the scattering, summed over $q_K=[0.4,0.6]$ rlu  for $B=0.025$ T (red) and 9 T (blue) at energy transfers (A) $0.5~\rm meV$ and (B) $1.0~\rm meV$. (C,D) The $q_L$ dependence of the scattering divided by the square of the form factor, summed over $q_K=[0.4,0.6]$ rlu for $B=0.025~T$ (red) and 9 T (blue) at energy transfers (C) $0.5~\rm meV$ and (D) $1.0~\rm meV$.  Error bars in all figures represent one standard deviation.}
\label{fig:fig4}
\end{figure}

\subsection*{Comparison of the Energy, Temperature, and Field Dependencies of $S\left(q_K\right)$}
To illustrate the dependencies of the raw data on energy, temperature, and field, we show cuts along $q_K$ computed using Eqn. S1, where the $q_{K}$ dependence of the scattering is related to the $d$-electron form factor as described above.  Fig. S5(A) shows the monitor normalized scattering intensities measured at $0.07~\rm K$  and a small field $0.025~\rm T$ for different energy transfers. It is clear that increasing energy transfer suppresses the scattering.  Fig. S5(B) shows the lack of any measurable temperature dependence of the scattering, which is measured in a magnetic field of $0.025~\rm T$ and an energy transfer of $0.35~\rm meV$.  Fig. S5(C) shows the magnetic field dependence of the scattering, measured at $0.07~\rm K$ and an energy transfer of $0.35~\rm meV$.  The sample was re-oriented between the data acquisitions in Figs. S5(A,B) and (C), leading to the slightly different lineshapes (see next section).

\begin{figure}
\centering
\includegraphics[width=.8\linewidth]{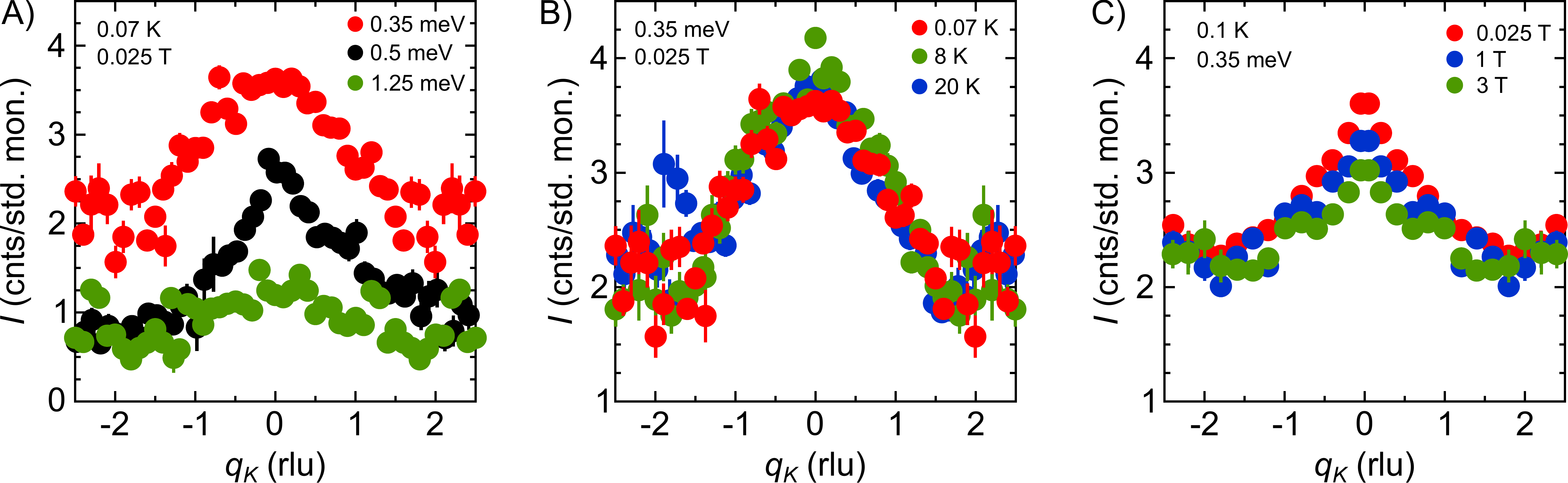}
\caption{{\bf $q_K$ dependence of the scattering at different energies, temperatures, and magnetic fields} (A) Cuts of $I(q_K)$ summed over $q_L = \left[0.8, 1.8\right]$ rlu at $T = 0.07$ K and fixed magnetic field $B = 0.025$ T for different energy transfers $E=0.35~\mathrm{meV}$ (red), 0.5 meV (black), and 1.25 meV (green). (B) Cuts of $I(q_K)$ summed over $q_L = \left[0.8, 1.8\right]$ rlu at a fixed energy transfer of $E=0.35$ meV and $B=0.025$ T for different temperatures $T=0.07$ K (red), 8 K (green), and 20 K (blue).  (C) Cuts of $I(q_K)$ summed over $q_L = \left[0.8, 1.8\right]$ rlu at fixed energy transfer $E=0.35$ meV and fixed temperature $T = 0.07$ K for different magnetic fields  $B = 0.025$  T (red), 1 T (blue), and 3 T (green).  Error bars in all figures represent one standard deviation.}
\label{fig:fig5}
\end{figure}

\subsection*{Comparison of Different Data Acquisitions}
The data presented in this paper were measured over the course of several experimental runs on MACS.  Between those times, several changes were made to the experimental conditions, involving changes in the filtering arrangements, magnet dark angle orientation, and the detector orientation with respect to the sample.  We made every effort to minimize other unintended and uncontrollable differences in the experimental conditions between the beam times, such as the exact size of the incident beam masks. We were able to accommodate all of these differences in experimental conditions by introducing a multiplicative constant that scaled the different data sets. This effect was accounted for in the conversion of the data into absolute units outlined in corresponding section of this supplement.

More drastically, however, before the final beam time in late May 2016, the sample was re-oriented into a new $a$-$c$ ($[H0L]$) scattering plane.  This was done to ensure that our MACS experiment was not simply seeing the tails from significant amount of scattering from an out of plane direction.  The results of that experiment were null and are not presented here.  When the sample was re-mounted in the original $b$-$c$ plane, one of the crystals became slightly misaligned out of the plane, changing the line shape of our $q_L$ integrated signal by a small amount.  This change only introduces a small difference in line shape,  and the field and temperature dependence are related to those found in the previous experiments through  a multiplicative factor.

Fig. S6 illustrates the effect of this procedure on the temperature dependence of $\chi ''$, measured for $B=0.025~\rm T$ and $E=0.35~\rm meV$ in both 2015 and 2016.  A  multiplicative factor to scale the integrated intensities maps the newer measurements directly onto the original measurements over two orders of magnitude in temperature for data measured in the same field and at the same energy.  The same multiplicative constant maps the field dependence at constant energy and temperature.

\begin{figure}
\centering
\includegraphics[width=.8\linewidth]{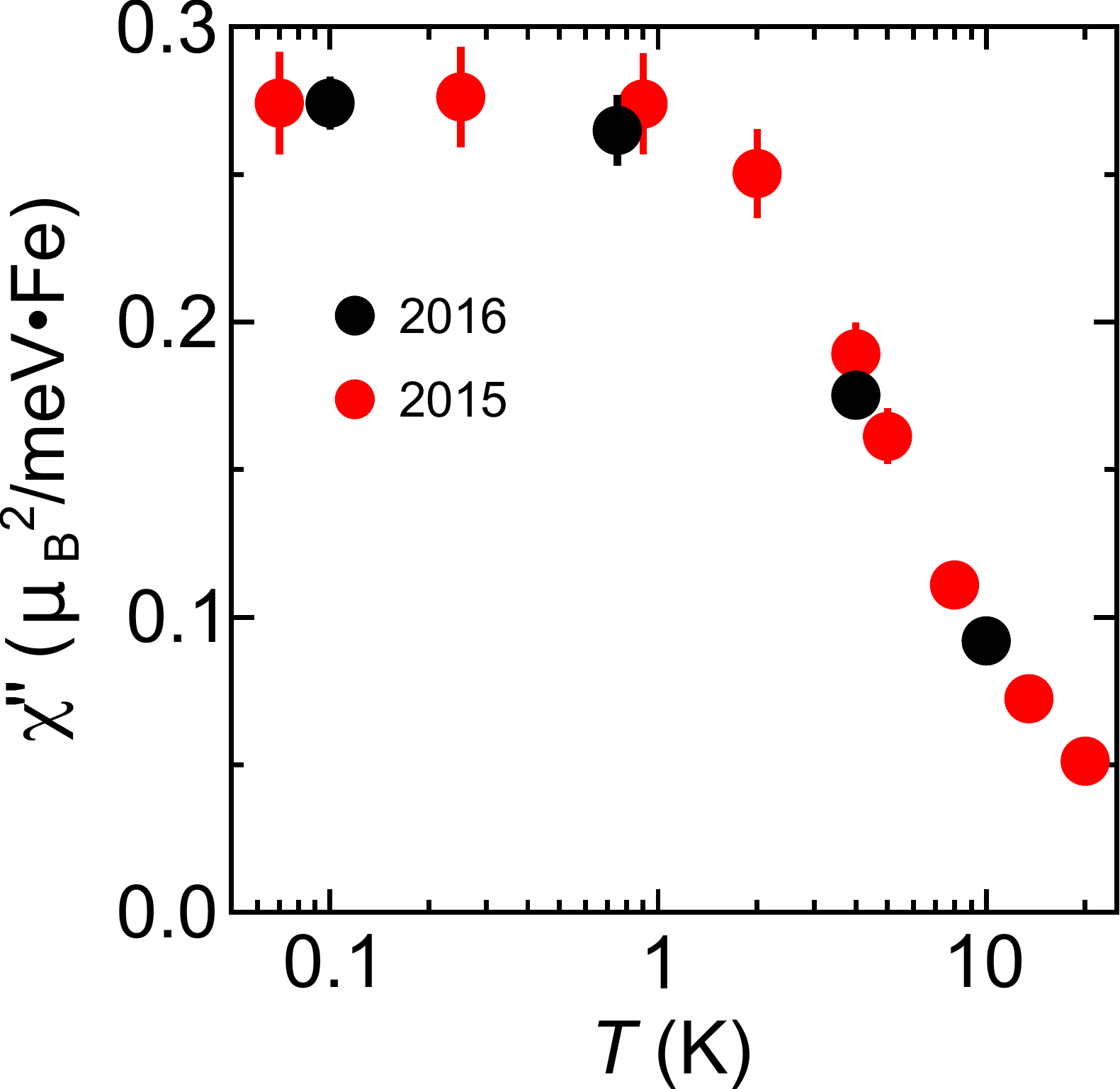}
\caption{{\bf Comparison of different data acquisitions.}  Data at $E=0.35$ meV and $B=0.025$ T analyzed as described in this supplement measured in 2015 (red), compared to data measured in 2016 (black) with a multiplicative scale factor to account for the difference in integrated intensity of the 2016 data.  Error bars in all figures represent one standard deviation.}
\label{fig:fig6}
\end{figure}

\subsection*{Temperature and Energy Dependent Cutoff of the Divergence of $\chi''$}

As noted in the main text discussion of Fig.~3, the exact form of energy and temperature dependent scaling of $\chi''$ that we observe implies a divergent $\chi''$ and magnetization that would violate the moment sum rule.  Our experimentally derived $E/T$ scaling must therefore necessarily be cutoff as $E, T\rightarrow 0$ to avoid this divergence.  While this is not a unique solution, a proposal that we tentatively put forward is the form $\chi'' \propto \left(E^2+\left(\pi k_BT\right)^2\right)^{-1.4/2}\tanh\left(E/k_BT\right)$.  This expression avoids this divergence problem and has the general properties required of our data in the limits of large and small values of $E/T$.    In Fig. S7(A) we present precisely the analysis shown in Fig. 3(D) of the main text, while scaling with the proposed cutoff form is shown in Fig.  S7(B).  Each form on a linear scale is respectively shown in Figs. S7(C,D) for comparison.  Fig. S7 demonstrates that this particular implementation of an energy and temperature cutoff for the dynamical susceptibility is generally consistent with our data, and we point out that agreement at large $E/k_BT$ could be improved by subtracting a small constant from $\chi''$ to isolate the purely quantum critical part.  Also note that the largest values of $E/k_BT$ are measurements made at the largest energy transfers, where our experimental signal to noise ratio is the smallest and where the uncertainty is correspondingly largest.  While this particular cutoff yields the expected limiting behaviors, a detailed test would require a more extensive data set for energies and temperatures giving large $E/k_BT$.  Given how small $\chi''$ becomes in this limit, it is likely that another experimental probe might be more suitable for such a test.

\begin{figure}
\centering
\includegraphics[width=.8\linewidth]{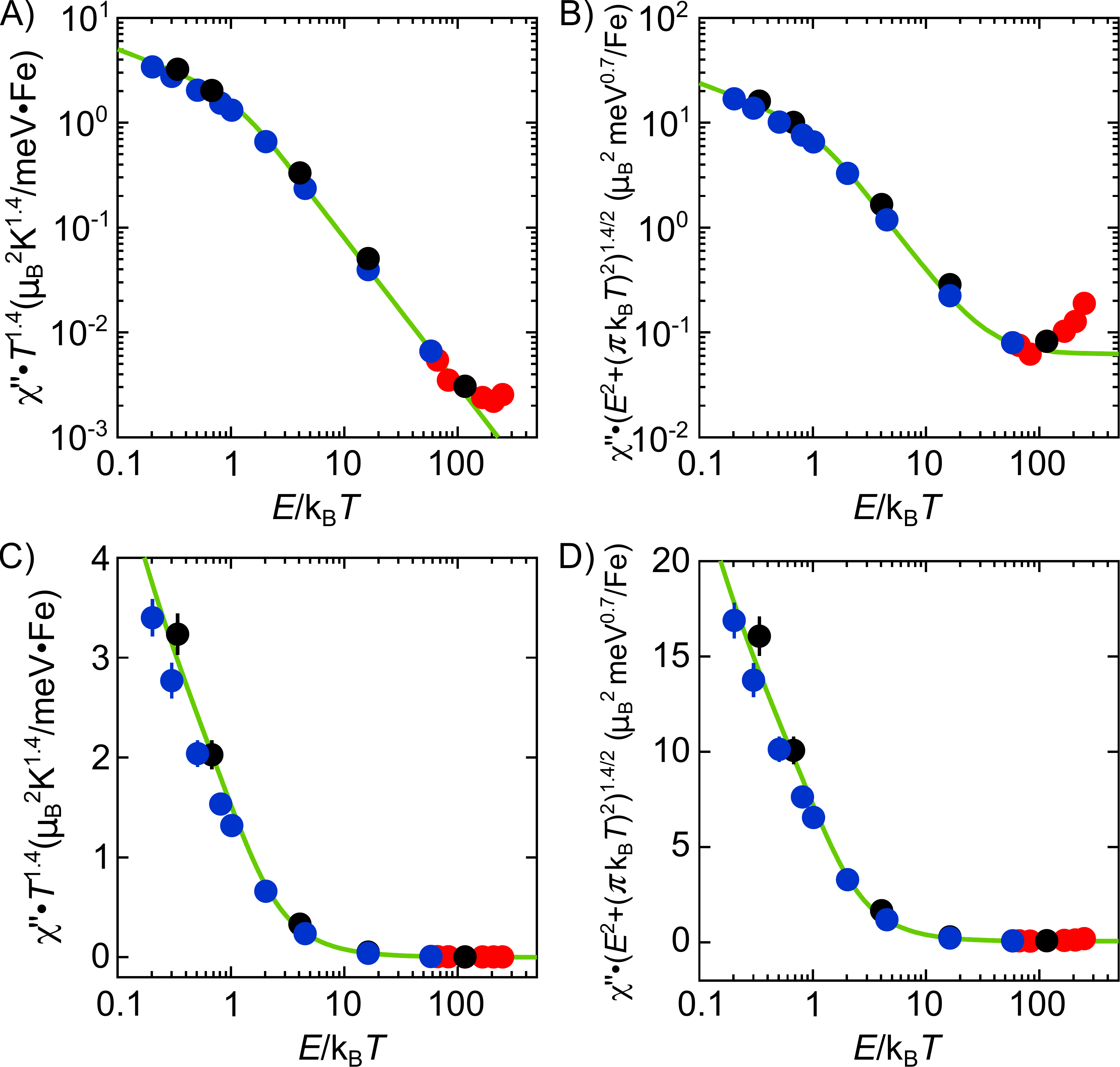}
\caption{{\bf Divergence in $E/T$ scaling.}  (A) The exact analysis as Fig.~3(D) of the main text, where $\chi''T^{1.4}$ measured at fixed energies $E=0.35~{\rm meV}$ (blue) and $E=0.7~{\rm meV}$ (black) and different energies at fixed $T=0.07~\rm K$ (red) is plotted as a function of $E/k_BT$.  The solid green line compares the scaled data to the function $\left(E/k_BT\right)^{-1.4}\tanh\left(E/k_BT\right)$.  (B) The same data as (A), but $\chi''$ is scaled by $\left(E^2+\left(\pi k_BT\right)^2\right)^{1.4/2}$.  Green line is $\tanh\left(E/k_BT\right)$, scaled by the same factor.  Panels (C) and (D) show the same data as panels (A) and (B) respectively, but on a linear vertical scale.}
\label{fig:fig7}

\end{figure}
\clearpage

\bibliographystyle{pnas-new}

\end{document}